\documentstyle[12pt]{article}

\begin{document}
\begin{titlepage}
\vspace*{-1cm}
\begin{flushright} RAL-94-104\\
DTP/94/78   \\
September  1994\\
\end{flushright}
\vskip 2cm
\begin{center}
{\Large{\bf Parton Distributions for low $Q^2$}}
\vskip 1.cm
{\large A.D.~Martin} and {\large W.J.~Stirling}
\vskip .2cm
{\it Department of Physics, University of Durham \\
Durham DH1 3LE, England }\\
\vskip .4cm
and
\vskip   .4cm
{\large R.G.~Roberts}
\vskip .2cm
{\it
Rutherford Appleton Laboratory,  \\
Chilton, Didcot  OX11 0QX, England
} \\
\vskip 1cm
\end{center}

\begin{abstract}
We extend the MRS(A) set of parton distributions,
 which provides an up-to-date description of  the structure
of the proton probed by hard interactions,  into the region of $Q^2$ below
$5\ {\rm GeV}^2$. After physically-motivated modifications, we obtain a smooth
description of the data on the structure
function $F_2(x,Q^2)$ all the way from $Q^2 = 10^{-1}$ to
$10^3\ {\rm GeV}^2$.
\end{abstract}
\end{titlepage}

\newpage

\section{Introduction}

Perturbative QCD is remarkably successful in describing a broad sweep of
hard-scattering processes involving the nucleon.
 This implies that it is possible to extract from
the data a consistent set of parton distributions  in the proton,
$f_i(x,Q^2)$, which
evolve in $Q^2$ according to the standard GLAP equations in the region where
perturbative QCD is appropriate.

Experimentally there has been steady progress,
first  in extending the kinematic
range and in improving the precision of the data (in particular of the
deep-inelastic structure function measurements) and, secondly, in extending
the range of ``hard" scattering data (e.g. the recent measurements
of the $W^\pm$ rapidity asymmetry \cite{CDFASYM} and of the asymmetry
in Drell-Yan production in $pp$ and $pn$ collisions \cite{NA51}).
 As a  result, modern sets of parton distributions \cite{MRSD,CTEQ,MRSA} are
strongly constrained. Here we concentrate on the most recent set of
partons, MRS(A) \cite{MRSA}, which were obtained in a global analysis
of the full range of deep-inelastic  and related data.
 These distributions  provide a detailed
description of the proton structure in the region where leading-twist
perturbative
QCD is valid, which may be taken as $Q^2 > Q^2_0 \approx  4\ {\rm GeV}^2$.  In
the
case of deep-inelastic scattering,
we know that higher-twist contributions become
important even above this value of $Q^2$, especially at large $x$, and
so an additional cut is always made in $W^2$, typically
$ W^2 > W^2_0 \simeq 10 \ {\rm GeV}^2$.

There are, however,  good reasons for requiring a
quantitative description of the nucleon structure at
lower values  of $Q^2$.  In structure function  measurements, for
example, the implementation of radiative corrections requires
 an approximate description of $F_2$ at low
$Q^2$.  Indeed, there now exist several low-$Q^2$ parametric forms for $F_2(x,
Q^2)$ \cite{BK,ALLM,NMCX,DL,CAP} which accurately describe $F_2$ in a
restricted region of
$Q^2$ and $x$.  An excellent summary of the situation has recently been
presented by Badelek and Kwiecinski \cite{BKREV}.  There is also a
demand for {\it individual} parton distributions
$f_i(x, Q^2)$ valid  at low $Q^2$ as well as for $Q^2> Q^2_0$.
  For example, many of the
measurements at HERA involve the use of
Monte Carlo event generators which describe
parton showering down to really low $Q^2$ values.

The aim of this paper is to provide a set of parton distributions that is
consistent with the data taken at low $Q^2$ (down to $Q^2 \simeq  0.1\ {\rm
GeV}^2$)
and, as $Q^2$ rises, smoothly approaches the set MRS(A) which is consistent
with
the high-$Q^2$ experimental data.  Note that there is already one set of parton
distributions, that of Gluck, Reya and
Vogt (GRV) \cite{GRV}, which could, in
principle, be used down to $Q^2$ = 0.3 GeV$^2$.
Their procedure is to start with a set of `valence'-like
quark and gluon distributions at $\mu^2 = 0.3\  {\rm GeV}^2$
and to evolve these up in $Q^2$
using the standard leading-twist ($= 2$) next-to-leading-order
evolution equations.  Above $Q^2 = O(2\ {\rm GeV}^2)$
these provide a reasonable description of $F_2$, consistent with the
belief that, by then, the
twist-two component is expected to dominate.  At low $Q^2$, however,
the higher-twist contributions are important
and the GRV partons alone do not give a good description of
$F_2$ there, as  was never their intention.
Moreover, there are also concerns that the perturbative GLAP
 evolution of valence-like GRV partons at small $x$ and  $Q^2$ is
unreliable \cite{EKL}.

In contrast, what we provide here are   `effective' low-$Q^2$ parton
distributions which are modified versions of the
leading-twist distributions, i.e.
they satisfy leading-twist $Q^2$ evolution only for
$Q^2 \lower .7ex\hbox{$\;\stackrel{\textstyle >}{\sim}\;$} 5 \ {\rm GeV}^2$.
In other words, the unknown higher-twist and non-perturbative components are
the origin of the empirical $Q^2$ modifications we impose on the distributions
at low $Q^2$, such that when these partons
are inserted into the formal expressions
for $F_2$,  we obtain detailed agreement with the  data.
Particularly useful in this context  are the new low-$Q^2$
structure function measurements  from the  E665 collaboration
\cite{E665}.

Note that the gluon and sea-quark distributions of the MRS(A) set of
partons  have the singular small-$x$ behaviour
\begin{equation}
xg, \; x q_{\rm sea} \sim x^{-0.3}
\label{eq:smallx}
\end{equation}
as $x\to 0$, which results from including the HERA measurements of $F_2$
in the global analysis. For such a singular gluon distribution, GLAP
evolution is expected \cite{EKL,M} to faithfully mimic the small-$x$
description of $F_2$ by the BFKL equation \cite{BFKL}.

\section{Parametrization of partons at low $Q^2$}

Our starting point is the MRS(A) set of parton distributions, which are
obtained from a global data analysis of deep-inelastic and related data
with $Q^2 > 5\ {\rm GeV}^2$ (and $W^2 > 10\ {\rm GeV}^2$).
The partons give an excellent description of these data. However our
goal here is to obtain partons which describe data to much lower values
of $Q^2$ (and $W^2$). We therefore begin by using the
next-to-leading-order GLAP equations to evolve down in $Q^2$ from the
MRS(A) starting distributions at $Q_0^2 = 4\ {\rm GeV}^2$.
At some stage we
anticipate that such a leading-twist GLAP-based extrapolation will fail.
Note that as we go below the charm threshold, the number of active
flavours in the evolution
drops from 4 to 3, and from our previous analysis \cite{MRSA} this threshold
is taken at $Q^2 = m_c^2 = 2.7\ {\rm GeV}^2$.

First we must allow for the effects of the target mass, which were not
included in the MRS(A) analysis. Such effects do not so much concern
low-$x$ data (where the photon-proton scattering energy is large) but
rather apply to data at large $x$ where they are significant even for
$Q^2 \lower .7ex\hbox{$\;\stackrel{\textstyle >}{\sim}\;$} 10\ {\rm GeV}^2$.
 Data in this kinematic region were excluded
from the MRS(A) analysis by the $W^2$ cut and so did not distort the
fit. To include the target mass corrections we replace $x$ by the target
mass variable $\xi$ \cite{DRGP}, where
\begin{equation}
\xi = \frac{2x}{1+r} \quad \mbox{with} \quad r^2 = 1+\frac{4M^2x^2}{Q^2}
\label{eq:target}
\end{equation}
where $M$ is the mass of the target. From Fig.~1 we see that this
straightforward correction gives a dramatic improvement  in the
description of the large $x$ SLAC measurements  of $F_2$ \cite{SLAC}.
Here the difference between the modified and unmodified data is almost
entirely due to the $x\to\xi$ substitution.

Our primary interest is the behaviour at low $Q^2$. There exist
fixed-target measurements of $F_2(x,Q^2)$ at low $Q^2$ and low $x$.
Indeed, the E665 collaboration have recently presented preliminary data
which access even lower $Q^2$ and $x$ values than hitherto.
We expect that $F_2(x,Q^2)$ reconstructed from MRS(A) partons evolved
to low $Q^2$ will, at some stage in the backwards evolution, begin to
overshoot the data, since the fundamental requirement that
\begin{equation}
F_2 \to O(Q^2)
\label{eq:photolimit}
\end{equation}
as $Q^2 \to 0$ is not embodied in the perturbative QCD evolution.
The most natural way to rectify this deficiency is to introduce a form
factor
\begin{equation}
\rho(x,Q^2) = {Q^2 \over Q^2 + m^2}
\label{eq:formfac}
\end{equation}
into the parton distributions, where the $x$ dependence of $m^2$ is to
be determined from the low-$Q^2$ data. In this way we achieve a
phenomenological description of higher-twist contributions  which may,
for example, arise from parton shadowing \cite{GLR}.  If the latter is the
dominant higher-twist effect, we would expect  $m^2$ to increase with
decreasing $x$, as can be seen from
 the following simple argument. Naively, we would expect shadowing
to become important when the total $gg$ `interaction area' becomes a
significant fraction of the transverse area ($\pi R^2$) in which the
gluons are confined within the proton. That is when
\begin{equation}
{n_g \sigma_{gg}  \over \pi R^2} \sim O(\alpha_s) \; ,
\label{eq:ssaturate}
\end{equation}
where $n_g \sim x g$ is the gluon density per unit rapidity.
Noting that $\sigma_{gg} \sim \alpha_s^2/Q^2$, this implies that the `critical
line'  in the $(Q^2,1/x)$ plot for the onset of shadowing has the form
\begin{equation}
Q^2 \sim m^2 \sim x^{-\lambda}
\label{eq:onset}
\end{equation}
where, for simplicity, we have assumed $xg, x\bar q \sim x^{-\lambda}$
and ignored the running of $\alpha_s$.

In practice it is clearly desirable  to distort the MRS(A) partons as
little as possible for $Q^2 > Q_0^2 \equiv 4\ {\rm GeV}^2$, and so we choose
\begin{equation}
m^2 = m_0^2(x)\; \exp(-Q^2/Q_0^2) \; .
\label{eq:msquared}
\end{equation}
Thus there is negligible modification to the partons
(i.e. $\rho \simeq 1$) in the kinematic region ($Q^2 > Q_0^2$) of the data that
originally
determined the MRS(A) distributions. The exponential factor in
(\ref{eq:msquared}) does not significantly distort the determination of
the $x$ dependence of $m^2$, since $m^2$ is mainly determined
by $F_2$ data with
$Q^2 \lower .7ex\hbox{$\;\stackrel{\textstyle <}{\sim}\;$} 1\ {\rm GeV}^2$.

In summary, we modify the evolved MRS(A) parton distributions so that
they take the form
\begin{equation}
f_i(x,Q^2) = \rho(x,Q^2)\; f_i^{\mbox{\tiny MRSA}}(\xi,Q^2) \; ,
\label{eq:fi}
\end{equation}
and determine the parameter $m_0^2$ in $\rho$ by fitting to the
$F_2(x,Q^2)$ data at low $Q^2$ for each value of $x$.
It is interesting to note that
the form factor modification and the $x\to\xi$ target mass correction
essentially decouple as they are relevant  in distinct kinematic
regimes, namely in low-$Q^2$, low-$x$ and in large-$x$, moderate-$Q^2$
regions respectively.

In principle it could be argued that a different form factor should be
used  for different parton distributions. However the low-$Q^2$, low-$x$
data are dominantly described by the sea-quark distributions. For
instance, we find no significant change in the results if we use an
unmodified distribution for the gluon, that is $\rho = 1$. (Computing
the value of $R=\sigma_L/\sigma_T$ using the QCD expression for $F_L$
evaluated with our modified distributions does not remove the
discrepancy between the SLAC data \cite{SLAC} and the QCD prediction.
 The values of
$R$ obtained by SLAC still tend to lie above the computed values.)

Finally we note a small technical point. When the MRS(A) parton
distributions are evolved down from $Q_0^2 = 4\ {\rm GeV}^2$, the gluon
distribution $xg(x,Q^2)$ begins to become negative over a small range of
moderate $x$, just below $Q^2 = 0.625\ {\rm GeV}^2$. We therefore freeze
the unmodified partons, $f_i^{\mbox{\tiny MRSA}}(x,Q^2)$,
for $Q^2 < 0.625\ {\rm GeV}^2$ at their values at $Q^2 = 0.625\ {\rm GeV}^2$,
i.e.
\begin{equation}
Q^2 \to Q^2 + (Q_c^2 - Q^2) \theta(Q_c^2 - Q^2)
\label{eq:freeze}
\end{equation}
in the argument of the $f_i^{\mbox{\tiny MRSA}}$, with $Q_c^2 = 0.625\
{\rm GeV}^2$.
The effect  on the behaviour of the $u$ sea-quark distribution is shown
in Fig.~2 for two relevant values of $x$. Since the unmodified
distributions are decreasing  only slowly  with $Q^2$ there is little
effect  on the description of $F_2$.

\section{The description of $F_2^p$ data at low $Q^2$}

The behaviour of the parameter $m^2(x)$ as a function of $x$
is determined by fitting to the $F_2$ data of E665 \cite{E665} and
SLAC \cite{SLAC},  and also the 90~GeV data
of NMC \cite{NMC}.  The NMC data for $Q^2 >$ 5 GeV$^2$ proved to be crucial
in constraining the MRS(A) partons. Although in principle the overall
normalization of the NMC data could be varied with
 respect to the other data sets, no correction was in fact needed
\cite{MRSA}.
When we include the E665 data in our analysis we  again
allow for an arbitrary relative normalization adjustment.
There is in fact a significant overlap in the $x, Q^2$ ranges of
 the NMC and E665
data sets and we find that consistency is achieved if the latter are
renormalized
up by 20\%. Although this may appear to be a large factor,
 the E665 data are still preliminary, with  a
figure of 10-20\% being quoted as  the typical systematic uncertainty
\cite{E665}.  Applying
this correction
factor 1.2 to the E665 data leads to a systematic $x$ dependence in
the values of $m_0^2(x)$
extracted from $F_2$ measurements of the three
 experiments, SLAC, NMC and E665.

The results are shown on a $\log m_0^2 - \log x$ plot in Fig.~3(a),
and indicate that a good fit to the $x$ dependence of $m_0^2$ can be
obtained by using the form
\begin{equation}
m_0^2(x) = A x^{-n} \; .
\label{eq:m1}
\end{equation}
We find $A = 0.07\ {\rm GeV}^2$ and $n=0.37$, corresponding to the straight
line in Fig.~3(a). It is remarkable that the $x^{-0.37}$ behaviour
obtained from low-$Q^2$ data is compatible with the observed
singular $x^{-0.3}$  gluon and
sea-quark small-$x$ behaviour, (\ref{eq:smallx}),
obtained from large $Q^2$ data.
This connection between independent results obtained from
 different kinematic regions
is suggested by the simple arguments which lead to (\ref{eq:onset}).
It is interesting to note that $m_0^2(x)$ could be
 equally well  by represented  by the alternative form
\begin{equation}
m_0^2(x) = A \exp \left[ \;B\sqrt{\ln \left(\frac{1}{x}\right)}\; \right] \; .
\label{eq:m2}
\end{equation}
as shown in Fig.~3(b). The straight line corresponds to
 $A=0.015\ {\rm GeV}^2$ and $B=1.54$ in (\ref{eq:m2}).
Note that the quoted values for $A$ and $n$ in (\ref{eq:m1})
are for fits are performed in the $\overline{\rm MS}$ scheme. A similar fit
 in the DIS factorization scheme gives  $A=0.055\ {\rm GeV}^2$ and
  $n=0.39$.

In Figs.~4 and 5 we show the description of the low-$Q^2$ measurements of
$F_2$ \cite{E665,SLAC,NMC} using the MRS(A) partons modified  as in
(\ref{eq:fi})  with $m_0^2$ given by (\ref{eq:m1}). In Fig.~6 we show,
at sample values of $x$, the continuation of the low-$Q^2$ description
to the recent HERA measurements of $F_2$ \cite{H1,ZEUS}.
 The continuous curves show that the partons give a good description of
$F_2$ throughout the range
$ 10^{-1} \lower .7ex\hbox{$\;\stackrel{\textstyle <}{\sim}\;$}
Q^2 \lower .7ex\hbox{$\;\stackrel{\textstyle <}{\sim}\;$} 10^3\
{\rm GeV}^2$. The predictions of the `dynamical' GRV partons \cite{GRV} are
also shown (by dotted curves) in Fig.~6. Within the scope of their
model, these partons provide an excellent description of $F_2$ for
$Q^2 \lower .7ex\hbox{$\;\stackrel{\textstyle >}{\sim}\;$} 1\ {\rm GeV}^2$
but below this value increasingly undershoot
the data. It can also be glimpsed from Fig.~6 that the recent HERA data
indicate that the GRV predictions increase a little too steeply
with decreasing $x$.

\section{$F_2^n$ at low $Q^2$}

There is also evidence for higher-twist corrections from SLAC and NMC
low-$Q^2$ data \cite{SLACn,NMCn} for the ratio of the neutron and proton
structure functions. The deviations from the MRS(A)
perturbative QCD predictions for $F_2^n/F_2^p$ occur in the interval
$x \sim 0.1$ to $0.3$  and so we introduce a further modification
to the {\it valence} quarks solely to describe these
deviations.\footnote{Note that the flavour-independent form factor
modification of Eq.~(\ref{eq:fi}) would leave the $F_2^n/F_2^p$
ratio essentially unchanged.}
We use the following simple one-parameter form which allows  $F_2^n$
to be varied at low $Q^2$ while leaving $F_2^p$ unaffected:
\begin{eqnarray}
u_v(x,Q^2) & \rightarrow & u^\prime_v (x,Q^2) = (1-\frac{r}{4}) u_v (x,Q^2) -
\frac{r}{4} d_v (x,Q^2) \nonumber \\
d_v(x,Q^2) & \rightarrow & d^\prime_v (x,Q^2) = (1+r) d_v (x,Q^2) +
r u_v (x,Q^2)
\label{eq:val}
\end{eqnarray}
 where
\begin{equation}
r = \frac{\tilde Q^2}{Q^2 + \tilde Q^2} \; .
\label{eq:theta}
\end{equation}

The $F_2^n/F_2^p$ data obtained by NMC are shown in Fig.~7. They have
been corrected for deuteron shadowing effects using the results of the
analysis of \cite{BKshad}.
We see that the predictions obtained from the MRS(A) partons (continuous
curves) tend to undershoot the data. This discrepancy, however, is much less
than for $F_2^p$ itself at low $Q^2$, as would be expected for the ratio
$F_2^n/F_2^p$ which depends primarily on the valence distributions.
The dashed curves, obtained using Eqs.~(\ref{eq:val}) and (\ref{eq:theta})
with $\tilde{Q}^2 = 0.12\ {\rm GeV}^2$, give a good description of the data.
Whether this simple parametrization is adequate for smaller values of
$Q^2$ is not certain. In any case, {\it nuclear} shadowing at such low $Q^2$
is expected to overwhelm such a simple partonic description.

In Fig.~8 we show the prediction of the modified MRS(A) partons
(continuous curves)
for the EMC(NA28) \cite{NA28} measurement of $F_2$ on a deuterium target.
In this case the curves (and not the data) have been corrected for deuterium
screening effects. The description is very satisfactory and mainly checks
the form factor modification of (\ref{eq:fi}). These data are not sufficiently
precise to further constrain the parameter $r$ in (\ref{eq:val}), since
the valence-quark contribution is small for those EMC data
which lie at low $Q^2$.
For comparison, we also  show the  GRV predictions for  $F_2^D$.
At $Q^2 = 0.35\ {\rm GeV}^2$ the GRV values are about a factor 2 below
the data, but very rapidly evolve upwards with increasing $Q^2$
to be in reasonable agreement with the data.

\section{Conclusions}

A low-$Q^2$ parton model is really a contradiction in terms.
At some stage as $Q^2$ decreases,
the description of the proton's structure cannot be expressed in terms of
single
parton densities with simple logarithmic behaviour in $Q^2$.
The contributions from
parton-parton correlation densities with power behaviour must enter and
eventually the non-perturbative behaviour dominates.  Moreover, all of these
contributions are expected to be process-dependent.
The set of distributions we have derived in this study correspond
 to what we may call `effective' partons, insofar
as power-law corrections like those introduced in Eq.~(\ref{eq:fi})
 would presumably arise from including
gluon-quark or quark-antiquark correlation distribution functions.
In the framework of the operator product expansion,
we assume that  these higher-twist terms
can be added to the leading-twist piece, each with their separate $Q^2$
dependence.  These effective partons do not of course
 satisfy the usual conservation laws --
number of valence quarks, total fractional momentum, etc. --
but the violation is below 20\% even for $Q^2 \sim 0.5\ {\rm GeV}^2$.
Their detailed structure at low $Q^2$ is also very dependent
on the precise form of the starting distributions at $Q_0^2 = 4\ {\rm GeV}^2$.
Note that, in contrast with the studies in Refs.~\cite{BK,ALLM,NMCX,DL,CAP},
we are not
attempting here to make a model which has a smooth transition
to the $Q^2 \to 0 $  photoproduction limit, where a parton-based approach
is clearly invalid.

{}From our study we may estimate, at least in the case
of `singular' MRS-type parametrizations,
 where the leading-twist GLAP evolution begins to become unreliable as
$Q^2$ decreases. Interestingly, this depends
on the value of $x$  in just such a way as if  the higher-twist
contribution arose from parton shadowing.
Some insight may be obtained by presenting the results as a function of $x$
at fixed $Q^2$. In Fig.~9 we display this dependence of the $F_2^p$ and
$F_2^D$ data for $Q^2 \simeq 0.3\ {\rm GeV}^2$. In this low-$Q^2$ regime the
data are reasonably flat, whereas we see that the unmodified
 MRS(A) partons (dashed curves)
partly retain the $x^{-0.3}$ behaviour of the distributions at the start
of the backward evolution. The higher-twist, form-factor modification
(continuous curves) restores the agreement with the data,
with the modification decreasing as $x$ increases.
 These data indicate large higher-twist
contributions, but as $Q^2$ increases towards $5\ {\rm GeV}^2$ these effects
rapidly decrease.

Backward evolution in $Q^2$ is much more
sensitive to the starting distributions than is forward evolution.
It is therefore possible to conceive `non-singular' starting
parametrizations (with a lower starting scale $Q_0$)
in which the higher-twist effects estimated in this way
are much smaller. In this case the rise in $F_2$
with decreasing $x$ seen at HERA is generated by the long evolution length.
The GRV partons \cite{GRV} are a parametrization of this type.
We show in Fig.~9  the GRV predictions, which at this value of
$Q^2$ undershoot the data with a shape which reflects their
valence-like input at $Q^2 = 0.3\ {\rm GeV}^2$.
With increasing $Q^2$, the agreement between the GRV description
and the data rapidly improves (see Fig.~6).

Our justification for discussing individual
parton distributions at very low $Q^2$
values is largely practical.  As mentioned earlier, parton showering
Monte-Carlo programs at HERA incorporate  parton densities at some
low cut-off in $Q^2$.  We are providing here  effective parton
densities which,  when substituted
into the next-to-leading-order expressions for $F_2(x,Q^2$), give
 a reliable description of the
data down to $Q^2 = 10^{-1}\ {\rm GeV}^2$ and, on the other hand, up to
$Q^2 = 10^3\ {\rm GeV}^2$.\footnote{The {\tt FORTRAN} code for the
distributions, in both the
$\overline{\rm MS}$ and DIS factorization schemes, can be obtained
by electronic mail from W.J.Stirling@durham.ac.uk}
Fig.~6 shows how $F_2 (x,Q^2$)
for small values of $x$ is well described throughout all of this region.

\section*{Acknowledgements}

We thank Jan Kwiecinski and Genya Levin for valuable discussions,
and Heidi Schellman for information concerning the E665 experiment.


\section*{Figure Captions}

\begin{itemize}

\item [{[1]}] Measurements of $F_2^p(x,Q^2)$ at large $x$ obtained by
SLAC \cite{SLAC} and BCDMS \cite{BCDMS}. The latter data are normalized
by a factor 0.98 as required by the global parton analysis. The broken
curves are obtained using MRS(A) partons \cite{MRSA}, and the continuous
curves using  MRS(A) partons modified as in (\ref{eq:fi}). For these
data the target mass correction (the substitution $x\to \xi$)
is the dominant modification.

\item [{[2]}] The sea-quark distribution $x \bar u(x,Q^2)$ at two values
of $x$. The upper curves correspond to MRS(A) partons, frozen at $Q^2 =
0.625\ {\rm GeV}^2$. The lower continuous curves correspond to MRS(A) partons
with the form factor modification as in (\ref{eq:formfac})
and (\ref{eq:fi}). Also shown
for comparison (dashed curves) are the predictions of the GRV set of
partons \cite{GRV}.

\item [{[3]}]  (a) The values of $m_0^2$ obtained from fitting to E665
\cite{E665}, SLAC \cite{SLAC} and NMC \cite{NMC}
measurements of
$F_2^p(x,Q^2)$ at different values of $x$ using MRS(A) partons modified
as in (\ref{eq:fi}). The straight line, $m_0^2 = 0.07 x^{-0.37}\
{\rm GeV}^2$, is the least-squares fit to the values of $m_0^2$. \\
(b) As for (a), but with $\log m_0^2 $ plotted as a function of
$\sqrt{\log(1/x)}$.  The straight line is the least squares fit,
 $m_0^2 = 0.015 \exp[ 1.54 \sqrt{\log(1/x)} ] \ {\rm GeV}^2$.

\item [{[4]}] The description of the E665 measurements \cite{E665} of
$F^p_2(x, Q^2)$ by the MRS(A) partons modified as in (\ref{eq:fi}).
The curves have been renormalized  downwards by a
factor 1.2 for the reasons described in the text.

\item [{[5]}] The description of the SLAC \cite{SLAC} and
 NMC \cite{NMC} measurements  of
$F^p_2(x, Q^2)$ by the MRS(A) partons modified as in (\ref{eq:fi}).

\item [{[6]}] Fixed-target \cite{E665,NMC} and HERA \cite{H1,ZEUS}
measurements of $F^p_2(x, Q^2)$ at selected values of $x$, as described
by MRS(A) partons modified as in (\ref{eq:fi}).
The predictions of the GRV parton
distributions \cite{GRV} are also shown (by the dotted curves)
for comparison.

\item [{[7]}] The description of the NMC measurements \cite{NMCn} of
$F^n_2/F^p_2$ by MRS(A) partons (continuous curves)
and by MRS(A) partons  modified as in (\ref{eq:fi})
 - (\ref{eq:theta}) (broken curves).

\item [{[8]}] The description of the EMC(NA28)  measurements \cite{NA28} of
$F^D_2$ by MRS(A) partons
 modified as in (\ref{eq:fi}) - (\ref{eq:theta}) (continuous curves).
The predictions of the GRV parton
distributions \cite{GRV} at very low $x$ are also shown (by the dashed curves)
for comparison. The statistical and systematic errors on the data
have been combined
in quadrature. There is an additional overall normalization uncertainty
of $\pm 7\%$.

\item [{[9]}] The description of the E665 $F_2^p$
(renormalized upwards by a factor 1.2) \cite{E665}
and  EMC(NA28) $F_2^D$   \cite{NA28}
\cite{E665} measurements at
fixed $Q^2 \simeq 0.3\ {\rm GeV}^2$
 by MRS(A) partons
 modified as in (\ref{eq:fi}) - (\ref{eq:theta}) (continuous curves).
The  unmodified MRS(A) predictions (dashed curves)  and
the predictions of the GRV parton
distributions \cite{GRV} (dotted curve)  are also shown.

\end{itemize}

\end{document}